\documentclass[3p,times]{elsarticle}

\usepackage{ecrc}

\volume{00}

\firstpage{1}

\journalname{Nuclear Physics A}

\runauth{A. Morsch}

\jid{nupha}

\usepackage{amssymb}

\usepackage[figuresright]{rotating}
\newcommand{\bq}{\begin{equation}}
\newcommand{\eq}{\end{equation}}

\newcommand{\bqq}{\begin{eqnarray}}
\newcommand{\eqq}{\end{eqnarray}}

\newcommand{\pt}{\ensuremath{p_{\rm{t}}}}
\newcommand{\jt}{\ensuremath{j_{\rm{t}}}}
\newcommand{\pta}{\ensuremath{p_{\rm t, assoc}}}
\newcommand{\ptt}{\ensuremath{p_{\rm t, trig}}}

\newcommand{\Dphi}{\Delta\varphi}
\newcommand{\Deta}{\Delta\eta}

\newcommand{\dd} {\mbox{${\rm d}$}}

\newcommand{\tev}     {\mbox{${\rm TeV}$}}
\newcommand{\gmom}    {\mbox{${\rm GeV}/c$}}

\begin{document}

\begin{frontmatter}



\dochead{}

\title{Jet-like near-side peak shapes in Pb--Pb collisions at $\sqrt{s_{\rm NN}} = 2.76 \, \tev $ with ALICE}

\author{Andreas Morsch \\
on behalf of the ALICE Collaboration}

\address{CERN, 1211 Geneva 23, Switzerland \\ andreas.morsch@cern.ch}

\begin{abstract}
We present preliminary results from di-hadron correlations in azimuth ($\varphi$)
and pseudo-rapidity ($\eta$) induced by hadronic di-jets produced in Pb--Pb collisions at 
$\sqrt{s_{\rm NN}} = 2.76 \, \tev$.  
The jet-like near-side peak has been isolated from $\eta$-independent collective flow correlations
using a $\eta$-gap method and its shape has been characterized by $rms$ and excess kurtosis. 
These quantities have been studied as a function of centrality as well as of trigger
(associated) particle transverse momentum ranging from $2-8 \, \gmom$ ($1-3 \, \gmom$).
We observe that the $rms$ in $\Deta$ increases with the centrality of the collisions whereas the 
$rms$ in $\Dphi$  stays constant within experimental uncertainties.
\end{abstract}


\end{frontmatter}


\section{Introduction}
High-$\pt$ partons traversing the high color-density medium produced in ultra-relativistic heavy ion collisions are expected
to lose energy due to induced gluon radiation and elastic collisions \cite{quenching}. 
There is overwhelming experimental evidence for partonic energy loss 
from measurements of the suppression of high-$\pt$ particles, back-to-back 
particle correlations and rates of fully reconstructed jets as well as from the energy imbalance 
of di-jet and $\gamma$-jet events \cite{evidence1, evidence2}.
At the LHC, the study of high energy jets by ATLAS and CMS has shown that the remnant jet 
has an unmodified structure and the radiated energy results in low-$\pt$ particles 
found far from the jet-axis \cite{evidence2}. This can be interpreted as a particular case of jet broadening in 
which the core region does not change and wide tails induced by gluon radiation develop. 
Furthermore, it has been predicted that the coupling to the longitudinally flowing medium \cite{armesto} 
or to turbulent color fields \cite{majumder}
will lead to broadening that is larger in $\eta$ than in $\varphi$ (excentric jets).
The motivation for the present analysis is to extend jet shape studies into the low and intermediate $\pt$ regions at the LHC,
where the broadening and flow effects are expected to be large.

\begin{figure}[ht]
\centering
\includegraphics[trim = 0 0 200 0, clip=true ]{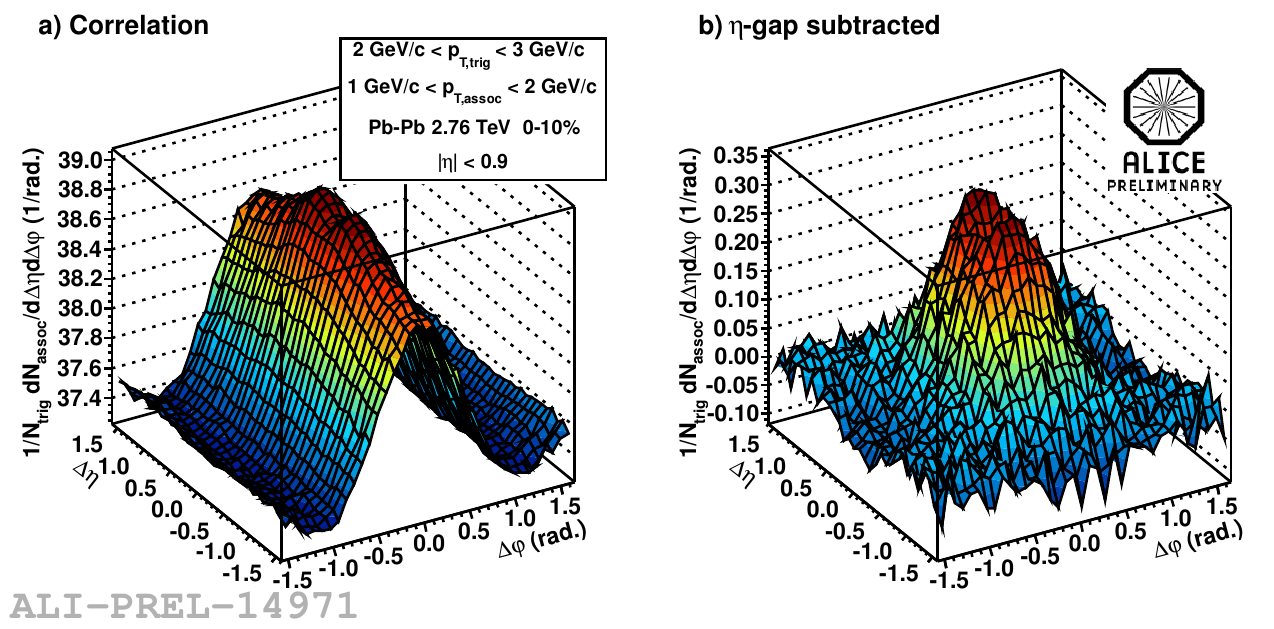}
\caption{
Per trigger associated yield before (a) and after (b) $\eta$-gap subtraction.
}
\label{fig:flowsubtraction}
\end{figure}

\section{Analysis Procedure}
At low and intermediate $\pt$ (from a few $\gmom$ up to about $20 \, \gmom$) it is impossible to reconstruct jets 
in heavy ion collisions over the fluctuations of the underlying event. 
However, jet properties can be studied also inclusively using di-hadron correlations.
To this end, one plots the angular differences $\Dphi$ and $\Deta$ between trigger particles 
and all associated particles satisfying pairs of cuts on ($\ptt, \, \pta$). 
The angular correlations are quantified by the per trigger associated yield $1/N_{\rm trig} \dd^2N/\dd\Dphi \dd\Deta$. 
In these one observes the so called near-side peak ($\Dphi = \Deta = 0$) and the away-side ridge 
($\Dphi = \pi$) induced by back-to-back parton production (Fig. \ref{fig:flowsubtraction} (a)).
In heavy ion collisions, collective flow results in an additional strong modulation in $\Dphi$ independent of $\Deta$.
In order to isolate the jet-like correlations, we use the region ($1 < |\Deta | < 1.6$), where flow dominates, to determine
and subtract flow correlations (Fig. \ref{fig:flowsubtraction} (b)). 
Since the $\Deta$ correlation between the trigger particle and the recoiling 
jet is very weak, the jet-related away-side ridge is also subtracted.  
Hence, one can not use this method to study the away-side region.

A detailed description of the ALICE experiment is given in Ref. \cite{alice}.
For the present analysis we used $1.5 \cdot 10^7$ events from the 2010 Pb--Pb data taking period and $5.5 \cdot 10^7$ events 
from the 2011 pp at $\sqrt{s} = 2.76 \, \tev$ run  as reference. The trigger and centrality selection for the Pb--Pb events 
are described in Ref. \cite{trigpb}. We select high-quality tracks within $|\eta | < 0.9$ reconstructed from Time Projection Chamber (TPC) clusters  
constraining the track parameters to the primary vertex. 
This leads to an optimal efficiency uniformity in azimuth while preserving a good $\pt$-resolution.
Two-track resolution effects have been minimized by cuts on the closest approach of tracks within the TPC volume.

To obtain the fully corrected per trigger associated primary particle yield,
a two step procedure is performed on the raw correlations. 
First we correct for two-track geometrical acceptance using the correlation function obtained from mixed events, i.e.
the trigger particle is taken from one event and the associated particle from a different one with similar centrality and 
longitudinal vertex position ($z$). In the second step the tracking efficiency correction and the correction for contamination from 
secondary particles is applied. 
The acceptance corrected correlation are calculated in each $z$-bin from the ratio of pair distributions from 
same ($N_{pair}^{same}(\Delta \varphi,\Delta \eta, z)$)
and mixed events 
($N_{pair}^{same}(\Delta \varphi,\Delta \eta, z)$).
\begin{equation}
\frac{\dd^{2} N^{raw}}{\dd\Delta \varphi \dd\Delta \eta} (\Delta \varphi, \Delta \eta, z) 
= \frac{1}{N_{trig}(z)} \frac{N_{pair}^{same}(\Delta \varphi,\Delta \eta, z)}{N_{pair}^{mixed}(\Delta \varphi,\Delta \eta, z)} \beta .
\end{equation}
The normalization factor $\beta$ is chosen such that the acceptance correction factor interpolated to 
($\Delta \varphi = \Delta \eta = 0$) gives unity. 
In the next step the $z$- and centrality (c) -bins are combined calculating the weighted average:
\begin{equation}
\frac{\dd^{2} N^{raw}}{\dd\Delta \varphi \dd\Delta \eta} (\Delta \varphi, \Delta \eta) 
= 
\frac{1}{\sum_{c} \sum_{z}{N_{trig}(c, z)}} \sum_{c} \sum_{z}{N_{trig}(c, z)\frac{\dd^{2} N^{raw}}{\dd\Delta \varphi \dd\Delta \eta} (\Delta \varphi, \Delta \eta, c,z)} .
\end{equation}

In order to quantify the peak shape we determine the $rms$ in the $\Deta$ and $\Dphi$ directions
($\sigma_{\Delta \eta}, \, \sigma_{\Delta \varphi}$)
as well as the peakedness using the so called excess kurtosis. The latter is defined as $\mu_4 / \mu_2^2 -3$,
where $\mu_2$ and $\mu_4$ are, respectively, the second and forth moment of the distribution. 
The offset (-3) adjusts the excess kurtosis such that it is 
0 for a Gaussian shape.
The $rms$ can be determined either from the projections or using a fit of the near-side peak to the sum of two Gaussian functions.
In the following we only present the results from the fitting procedure.
\begin{figure}[ht]
\begin{minipage}[b]{0.5\linewidth}
\centering
\includegraphics[width = 1.00\linewidth]{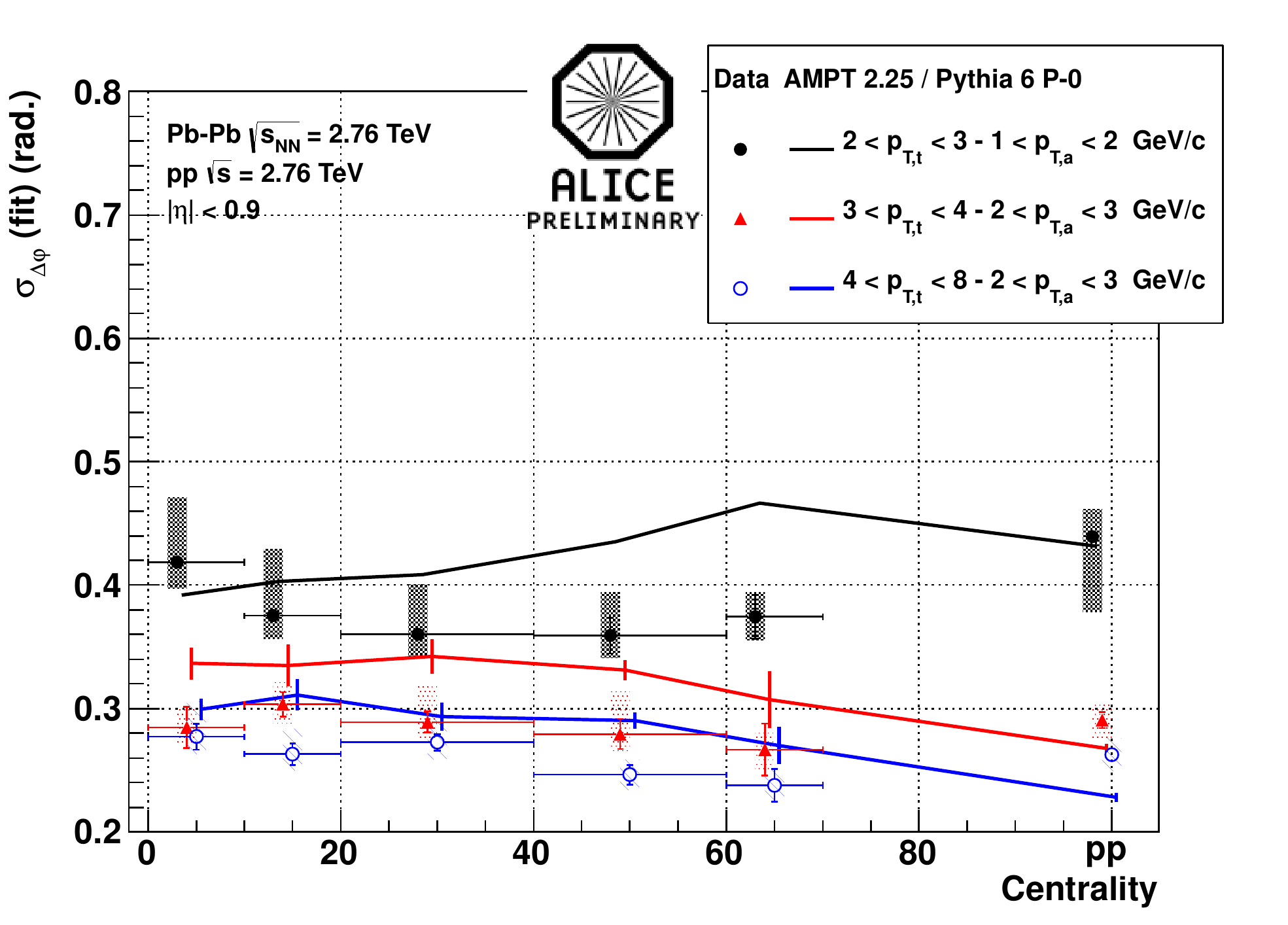}
\end{minipage}
\begin{minipage}[b]{0.5\linewidth}
\centering
\includegraphics[width = 1.00\linewidth]{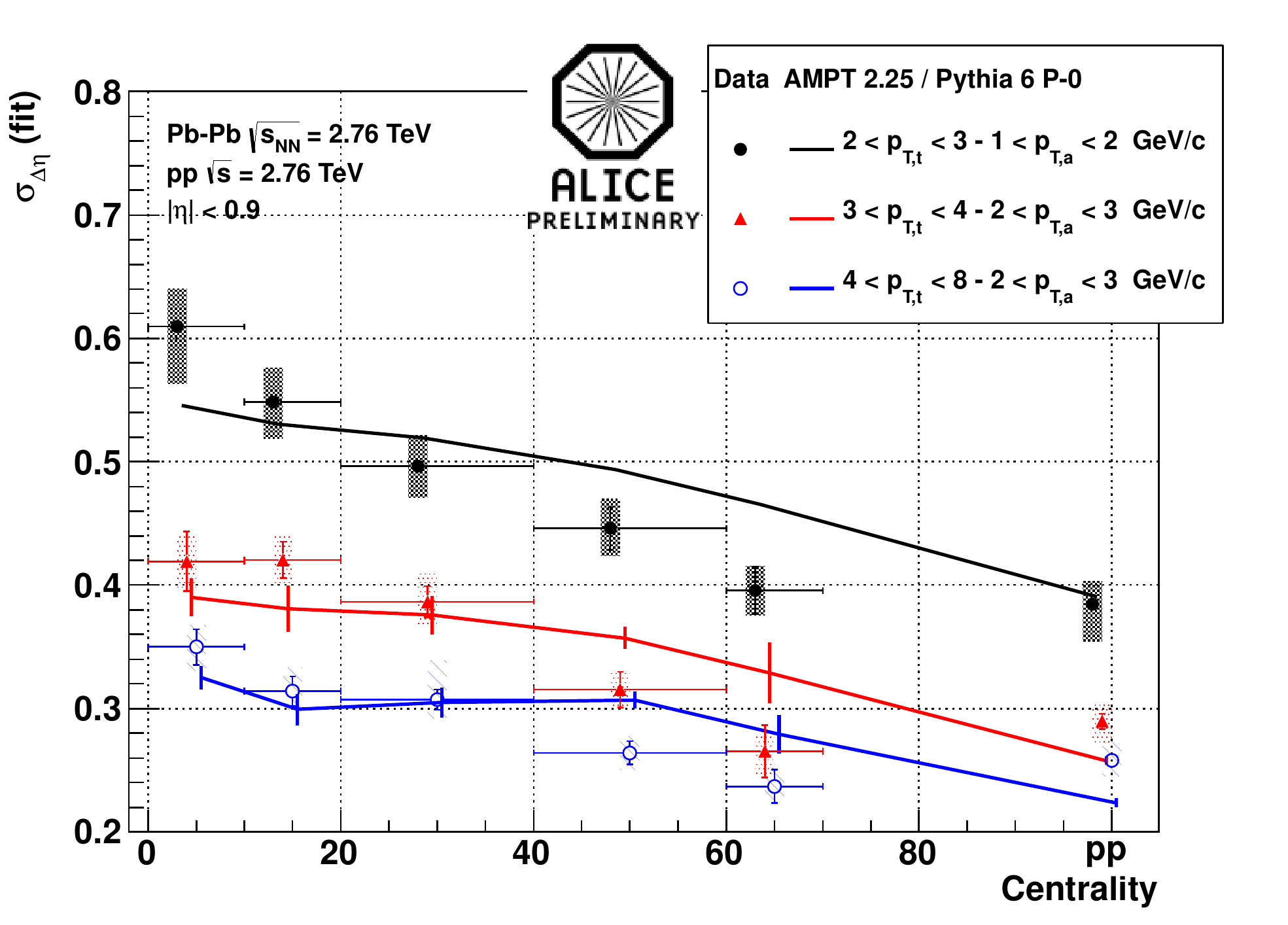}
\end{minipage}
\caption{Centrality dependence of $\sigma_{\Delta \varphi}$ (left) and $\sigma_{\Delta \eta}$ (right)
for three different bins of $\ptt$ and $\pta$. The lines are from the results of AMPT \cite{ampt} (Pb--Pb) and 
Pythia \cite{pythia} simulations (pp).}
\label{fig:res1}
\end{figure}

\begin{figure}[ht]
\begin{minipage}[b]{0.5\linewidth}
\centering
\includegraphics[width = 1.00\linewidth]{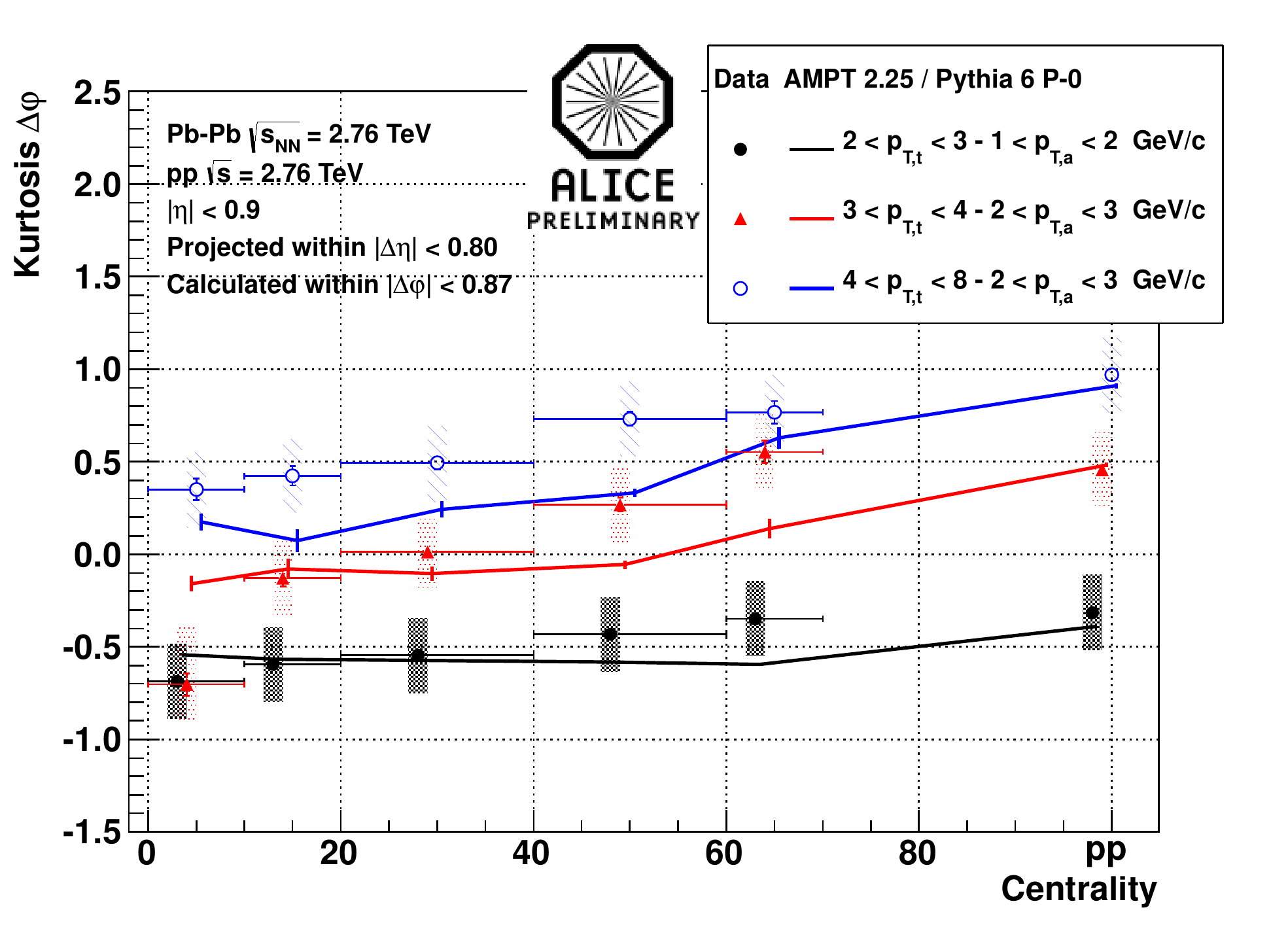}
\end{minipage}
\begin{minipage}[b]{0.5\linewidth}
\centering
\includegraphics[width = 1.00\linewidth]{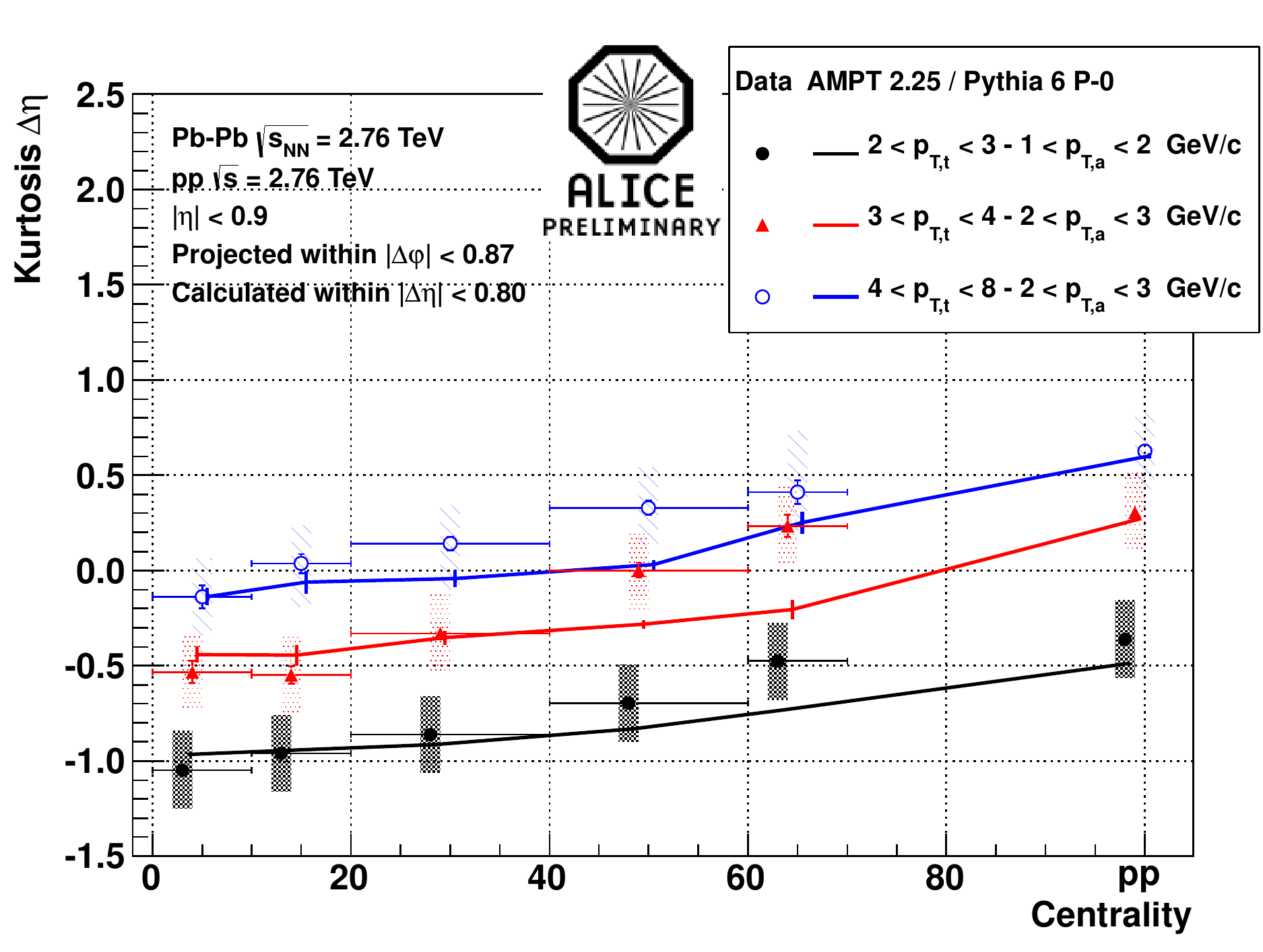}
\end{minipage}
\caption{Centrality dependence of the excess kurtosis in $\Dphi$ (left) and $\Deta$ (right)
for three different bins of $\ptt$ and $\pta$.
The lines are from the results of AMPT \cite{ampt} (Pb--Pb) and 
Pythia \cite{pythia} simulations (pp).}
\label{fig:res2}
\end{figure}
\section{Results}
Fig. \ref{fig:res1} shows the centrality dependence of $\sigma_{\Delta \eta}$ and $\sigma_{\Delta \varphi}$
for three different bins of $\ptt$ and $\pta$. The centrality $100\%$ data points are the pp reference.
No centrality dependence of $\sigma_{\Delta \varphi}$ is observed within errors, whereas 
there is a significant increase of $\sigma_{\Delta \eta}$ moving from pp to central collisions.
The $rms$ of both projections shows a strong $\pt$ dependence, mainly on $\pta$. This results from the 
fact that jet fragmentation is governed by a typical average momentum perpendicular to the jet axis 
($\jt \approx \sigma_{\Delta \eta , \Delta \varphi} \pta$) .

The corresponding dependencies of the excess kurtosis in the $\Deta$ and $\Dphi$ projections are shown in Fig. \ref{fig:res2}.
In all cases the excess kurtosis (peakedness) decreases moving from pp to peripheral and to central collisions
and increases with increasing ($\ptt , \pta$).

The data points are compared with the result of a AMPT 2.25 (with string melting) Monte Carlo simulation \cite{ampt} shown as lines.
The pp values have been simulated with Pythia 6.4 (Perugia-0) \cite{pythia}.
AMPT simulates the initial conditions using HIJING \cite{hijing}. It includes re-scattering of partons and hadrons.
Hadronization is simulated using the Lund model and coalescence. 
The AMPT simulations are in good agreement with our measurements. More detailed studies are needed to find out which of the 
features implemented in AMPT are responsible for the $\Deta$ broadening.

\begin{figure}[ht]
\begin{minipage}[b]{1.0\linewidth}
\centering
\includegraphics[bb = 190 0 550 180, clip]{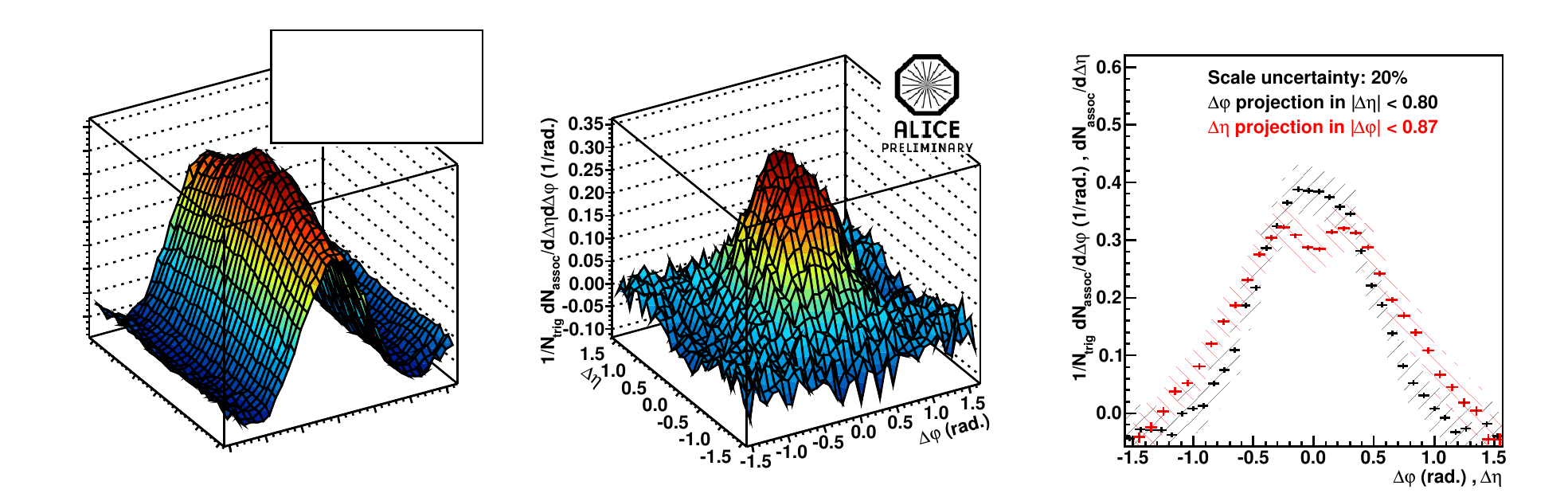}
\end{minipage}
\begin{minipage}[b]{1.0\linewidth}
\centering
\includegraphics[trim = 370 0 0 0, clip=true, scale=1.00]{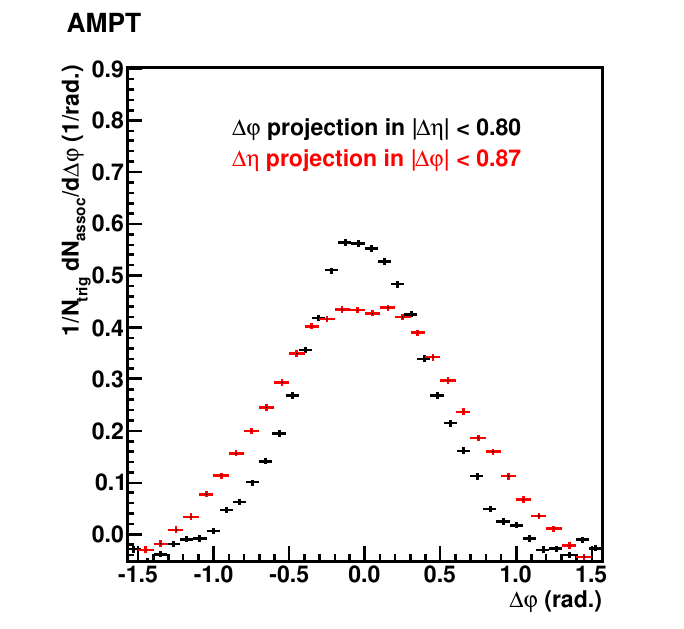}
\end{minipage}
\caption{Per trigger associated yield differential in $\Dphi$ and $\Deta$ for the $10\%$ most central collisions and 
the lowest $\pt$ bin studied ($2 < \ptt < 3 \, \gmom$, $1 < \pta < 2 \, \gmom $)(upper left). Corresponding projections (upper right).
Corresponsing projections from AMPT model simulations (lower).}
\label{fig:peak}
\end{figure}

Let us now have a closer look at the near-side peak for the $10\%$ most central collisions and 
the lowest $\pt$ bin studied ($2 < \ptt < 3 \, \gmom$, $1 < \pta < 2 \, \gmom $) for which the 
$\Deta$ width is largest (Fig. \ref{fig:peak}). This distribution shows a new characteristics; it starts 
to flatten in the region $|\Deta| < 0.4$. Also this behavior is reproduced by the AMPT simulation.





\bibliographystyle{elsarticle-num}
\bibliography{<your-bib-database>}







\end{document}